\documentclass[aps,pra,twocolumn,groupedaddress,showpacs,floatfix,superscriptaddress]{revtex4}
\usepackage{amssymb}
\usepackage{graphicx}
\usepackage{subfigure}	 	
\usepackage[english]{babel}
\usepackage{float}
\usepackage{amsmath}
\usepackage{calc}
\usepackage{bm}
\usepackage{color}
\usepackage{braket}
\usepackage{dcolumn}

\begin{document}

\title{Effect of laser phase noise on the fidelity of optomechanical quantum memory}

\author{F. Farman and A. R. Bahrampour}
\affiliation{Department of Physics, Sharif University of Technology, Tehran, Iran}

\begin{abstract}
Optomechanical and electromechanical cavities have been widely used in quantum memories and quantum transducers.
We theoretically investigate the robustness of opto(electro)-mechanical quantum memories against the noise of the control laser.
By solving the Langevin equations and using the covariance matrix formalism in the presence of laser noise, the storing fidelity of Gaussian states is obtained. It is shown that, the destructive effect of phase noise is more significant in higher values of coupling laser amplitude and optomechanical coupling strength $ G $. However, by further increasing coupling coefficient, the interaction time between photons and phonons decreases below the coherence time of laser frequency noise and the destructive effect of laser phase noise on the storing fidelity drops as well.
\end{abstract}

\pacs{42.50.Lc, 42.50.Wk, 42.50.Pq, 42.50.Ex}

\maketitle
\section{Introduction}
Optical quantum memories are one of the most important and essential equipments in quantum computers and quantum repeaters \cite{Lvovsky1}. Several methods for preparation of these memories have been developed over the last few decades \cite{Lvovsky1, Hosseini2, England3, Sparkes4}. One of the most famous method is based on the Electromagnetically Induced Transparency (EIT) effect, which has been observed in lambda-shaped atoms \cite{Harris5, Fleischhauer6, Lukin7,Mucke8}. Using this effect, by changing the coupling coefficient between light and atom, the group velocity of light can approach zero which results in mapping quantum state of light to an ensemble of atoms, quantum state \cite{Fleischhauer9,Fleischhauer10,Phillips11}. The upper limit of the storage time is of the order of the decoherence time of the atom \cite{Phillips11}.
In recent years optomechanical systems have had many applications specially as a macroscopic lab for observing quantum mechanical effects. By adjusting the input laser detuning, the mechanical mode of these systems can be cooled close to the ground state \cite{Gigan12, Wilson-Rae13, Schliesser14, Chan15}.
Recently, the EIT like effect has been observed in optomechanical cavities which is called optomechanically induced transparency (OMIT) \cite{Weis16, Agarwal17, Huang18}. OMIT can provide a new approach to designing quantum delay lines and memories \cite{Jiang19, Safavi-Naeini20}. Quantum state conversion between optical and mechanical modes of cavity has also been studied actively \cite{Zeng21, Zhang22, Tian23, Palomaki24}.
It has been shown that, in the strong optomechanical coupling regime of operation, if the optomechanical coupling is quantum coherent, the optical and mechanical modes are exchanged with the frequency proportional to the optomechanical coupling coefficient \cite{Tian23, Wang25, Verhagen26}.
This effect can be applied in the design of optical quantum memories and transducers which transfer the quantum state of light from one frequency to the other \cite{Wang25, Tian27}.
The storage time of optomechanical quantum memories is limited to $ \frac{1}{\gamma} $ where $ \gamma $ is the mechanical decay rate of the oscillator.  Optomechanical memories and transducers have been studied theoretically as well as experimentally in many groups \cite{Wang25, Fiore28, McGee29, Wang30}. In reference \cite{McGee29}, the storing fidelity has been obtained based on the quantum state diffusion method (QSD). As it is clear, the noise from the control (coupling) laser has a damaging effect on the storing fidelity.
Fundamentally, the intensity and phase noises of the control laser are related by the uncertainty principle. In our analysis it is assumed that by stabilizing the laser intensity, the intensity noise is negligible and phase noise becomes the main component of the control laser noise.
The effects of laser phase noise in cooling rate of mechanical mode and entanglement, have been studied by many researchers \cite{Rabl31, Ghobadi32, Yin33, Phelps34, Farman35, Abdi36}.
According to Rabl et al. if we consider the laser phase noise to be white noise, we won't be able to cool the the mechanical resonator to reach the ground state. However, experiments have shown the ground state cooling to be a possibility. Due to the finite coherence time of laser frequency noise, the phase noise must be considered as a color noise with finite bandwidth \cite{Rabl31, Diosi37}.
The phase noise spectral density is equivalent to the noise spectral density of a low pass filter excited by white noise \cite{Scully38}.
The cutoff frequency of the low pass filter is inversely proportional to the coherence time of the laser frequency noise. The flat spectral density, corresponds to the zero coherence time of input frequency noise (white noise). In this paper the effect of control laser phase noise on the storing fidelity of a Gaussian quantum state in the mechanical mode of the cavity, is investigated.
It is shown that, the quantum memory fidelity destroy drastically in the presence of white phase noise. The effect of laser frequency noise coherence time on memory fidelity is also taken into consideration.
In order to achieve the strong optomechanical coupling, the microwave electromechanical system that are easily cooled to temperatures below $ 100\ mK $ \cite{Teufel39} and an optomechanical Fabry- Perot system, are considered \cite{Groblacher40}.
Effects of optomechanical coupling strength on memory fidelity in the presence and in the absence of laser phase noise, is studied. It is shown that by raising the intracavity amplitude, the destructive effect of phase noise increases.
This paper is organized as follows: in section II, the theoretical model of the system is explained and the dynamics of the variables considering the effect of phase noise, are obtained. The noise model is also introduced in this section. The method of obtaining the storing fidelity of a Gaussian state and the storage protocol is described in Section III. Section IV is devoted to the numerical results and analysis of the effect of phase noise on the storing fidelity of electromechanical and optomechanical systems, and finally the paper is concluded in Section V.

\section{Theoretical Model}
We consider an optomechanical system for storing the quantum state of light. This system can be electromagnetic cavities such as Fabry-Perot \cite{Groblacher40}, microsphere or microtoroid cavities with one mechanical mode \cite{Verhagen26}, or electromechanical systems based on microwave resonance circuit \cite{Teufel39}. The schematics of a Fabry-Perot cavity with one movable mirror and a microwave circuit are shown in Figs. 1(a) and 1(b), respectively.
\begin{figure}[htbp]
\centerline{\includegraphics[width=0.7\columnwidth]{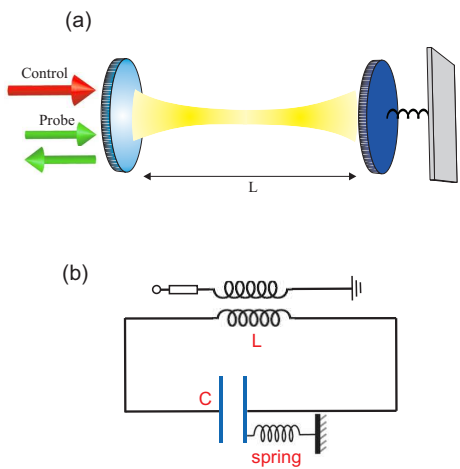}}
  \caption{(color online)
  Schematics of the following systems: a) An optomechanical Fabry-perot cavity consisting one vibrating mirror. The cavity is driven by a laser from the left mirror. b) A microwave cavity.}
\end{figure}
The cavity is driven by an intense coupling laser beam with frequency  $ \omega_L $,  which is red detuned from the cavity resonance frequency $ \omega_C $. The radiation pressure force of the driving field leads to displacement of the movable mirror, changing the resonance frequency of the cavity and coupling the intracavity field to the mechanical mode of the mirror. In microwave cavity, the displacement of the movable plate of capacitor due to the radiation pressure force of the microwave field, changes the capacitance $ C $ and resonance frequency of the microwave circuit ($ \omega_C =\frac{1}{\sqrt{LC}}$), where $ L $ is the inductance of the microwave resonator \cite{Teufel39}. High quality factor microwave resonators can be reached by employing super conductors.
In the resolved sideband limit (when the mechanical resonance frequency $ \omega_{m} $ exceeds the cavity decay rate $ \kappa $), if the coupling laser is tuned to the lower optomechanical sideband, in the rotating wave approximation (RWA), the linearised interaction Hamiltonian between mechanical and optical mode of the cavity is $  H_{int}=-\hbar G (\hat{a}^{\dagger}   \hat{b}+  \hat{a} \hat{b}^{\dagger})$ which is beam-splitter-like Hamiltonian \cite{Tian23, Wang25, Verhagen26}.  $ \hat{a} $ and $ \hat{b} $ are the linearised bosonic annihilation operators for cavity optical photon and mechanical phonons, respectively. $ G $ is the coupling coefficient between the optical and mechanical modes, which is proportional to the driving field. In the absence of any optical and mechanical dissipation, $ \hat{a}(t) $ and $ \hat{b}(t) $ can be derived easily:
\begin{eqnarray}\label{Eq.1}
\hat{b}(t)&=&\hat{b}(0) cos(Gt)+i \hat{a}(0) sin(Gt)
\nonumber \\
\hat{a}(t)&=&\hat{a}(0) cos(Gt)+i \hat{b}(0) sin(Gt)
\end{eqnarray}
As it can be seen from Eq. (\ref{Eq.1}), the quantum state of the mechanical and optical modes can transform into one another (up to a phase factor) with period $ \frac{\pi}{2G} $. This conversion is the basis of the design of quantum memory by opto(electro)-mechanical systems.
Initially we should prepare the optical mode of the cavity in the state $ \ket{\psi} $ we wish to store it in. By using the optomechanical coupling $ \frac{\pi}{2} $ pulse ($ \Delta  t=\frac{\pi}{2G}$), the optical state of interest is mapped to the mechanical mode. After the storage time $ \tau $ is up, the second optomechanical coupling $ \frac{\pi}{2} $ pulse will map the state that is stored in the mechanical mode to optical mode and we can retrieve it \cite{Tian23, Wang25}. In practice due to the existence of optical and mechanical energy decay rates and input noises, the initial pure state converts to mixed state and the fidelity of the final state decreases. In order to have a quantum memory, the optomechanical coupling must be larger than both the optical decay rate $ \kappa $ and the thermal decoherence rate of the system $  \gamma_{th}=N_{m} \gamma $ ($\gamma_{th} ,\kappa< G $), where $ N_{m} $ and $ \gamma $ are the mean thermal phonon number of reservoir and mechanical damping rate of the oscillator, respectively. If this condition is satisfied the optomechanical coupling is quantum coherent \cite{Verhagen26}. Our study is concentrated on the effect of the driving laser phase noise on the retrieved state fidelity in the quantum coherent regime of operation.

The Hamiltonian of the system in a frame rotating at the driving laser frequency can be written as:
\begin{eqnarray}\label{Eq.2}
\hat{H}&=&\hbar \omega_{m} \hat{b}^{\dagger} \hat{b}+\hbar \Delta_{0}  \hat{a}^{\dagger} \hat{a} -\hbar g_{0} \hat{a}^{\dagger} \hat{a}( \hat{b}+  \hat{b}^{\dagger})
\nonumber \\
&+&i \hbar E_{L}(t) (e^{i \varphi(t)}\hat{a}^{\dagger}-e^{-i \varphi(t)}\hat{a})
\end{eqnarray}
where $ \Delta_{0}=\omega_C-\omega_L $ is the detuning of coupling laser from the cavity resonance frequency. $ \varphi(t) $ is the fluctuating phase noise of the driving laser. $ E_{L}(t)=E_{L}+\delta E_{L}(t) $ is the complex amplitude of the driving laser inside the cavity which is composed of one deterministic term ($ E_{L} $) and one fluctuating term ($ \delta E_{L}(t) $) which arises from fluctuations in the amplitude of the laser. The deterministic term is related to the input power $ P $ from $ E_{L}=\sqrt{\frac{\kappa P}{\hbar\omega_L }} $. In a stabilized laser source, the amplitude noise of the driving laser is negligible compared to the phase noise, so we neglect the amplitude noise of the laser in this paper \cite{Abdi36}. $  g_{0}=-\frac{\partial \omega_C}{\partial L}x_{zpf} $ is the single photon optomechanical coupling  strength where $ L $ is the length of the cavity and $ x_{zpf}=\sqrt{\frac{\hbar}{2m\omega_m}} $ is the zero-point fluctuation amplitude of the mechanical oscillator, with $ m $ being the effective mass of the mechanical mode.

\subsection{Quantum Langevin equations}
Quantum Langevin equations of the system can be derived by adding the corresponding noise and damping terms to the Heisenberg equations:
\begin{eqnarray}\label{Eq.3}
\frac{d \hat{a}}{d t}&=& -i \Delta_{0} \hat{a}+i g_{0} \hat{a}(\hat{b}+\hat{b}^{\dagger})+E_{L}e^{i \varphi(t)}-\frac{\kappa}{2}\hat{a}+\sqrt{\kappa} \hat{a}_{in}
\nonumber \\
\frac{d \hat{b}}{d t}&=& -i \omega_{m} \hat{b}+i g_{0} \hat{a}^{\dagger} \hat{a}-\frac{\gamma}{2}\hat{b}+\sqrt{\gamma} \hat{b}_{in}
\end{eqnarray}
where $ \hat{a}_{in} $ and $ \hat{b}_{in} $ are input vacuum and thermal noise operators with two times correlation functions given by \cite{Zhang22, Gardiner41}:
\begin{eqnarray}\label{Eq.4}
 \langle \hat{a}_{in}(t) \hat{a}_{in}^{\dagger}(\acute{t})\rangle &=& (1+N_{C})\delta (t-\acute{t})
  \nonumber \\
 \langle \hat{a}_{in}^{\dagger}(t)\hat{a}_{in}(\acute{t}) \rangle &=& N_{C}\delta (t-\acute{t})
 \nonumber \\
 \langle \hat{b}_{in}(t) \hat{b}_{in}^{\dagger}(\acute{t})\rangle &=& (1+N_{m})\delta (t-\acute{t})
 \nonumber \\
 \langle \hat{b}_{in}^{\dagger}(t) \hat{b}_{in}(\acute{t})\rangle &=& N_{m}\delta (t-\acute{t})
\end{eqnarray}
Here $ N_{m(C)}=\frac{1}{e^{\frac{\hbar \omega_{m(C)}}{K_{B}T}}-1} $, is the mean thermal phonon(photon) number of the thermal(optical) reservoir in the frequencies $ \omega_{m}(\omega_{C}) $ at temperature $ T $ and $ K_{B} $ is Boltzmann's constant. At optical frequencies $  N_{C}\simeq 0 $. The statistical properties of phase noise will be determined later.
Using the transformation $ \hat{a}\rightarrow   \hat{\tilde{a}}e^{i \varphi (t)}$, we will be in the rotating frame with instantaneous fluctuating frequency $ \dot{\varphi}(t) $ \cite{Abdi36}.
\begin{eqnarray}\label{Eq.5}
\frac{d \hat{\tilde{a}}}{d t}&=& -i \Delta_{0} \hat{\tilde{a}}-i \dot{\varphi}\hat{\tilde{a}}+i g_{0}\hat{\tilde{a}}(\hat{b}+\hat{b}^{\dagger})+E_{L}-\frac{\kappa}{2}\hat{\tilde{a}}+\sqrt{\kappa} \hat{\tilde{a}}_{in}
\nonumber \\
\frac{d \hat{b}}{d t}&=& -i \omega_{m} \hat{b}+i g_{0} \hat{\tilde{a}}^{\dagger} \hat{\tilde{a}}-\frac{\gamma}{2}\hat{b}+\sqrt{\gamma} \hat{b}_{in}
\end{eqnarray}
The standard method is used for linearising the non-linear equations (\ref{Eq.5}). Operators are written as the sum of steady state values and fluctuating terms with zero mean value ($ \hat{\tilde{a}}\rightarrow \alpha_{s}+\delta \hat{a}, \hat{b}\rightarrow \beta+\delta \hat{b} $). The steady state values (fixed points) are obtained by setting the time derivatives of operators to zero ($ \alpha_{s}=\frac{E_{L}}{i(\Delta_{0}-g_{0}(\beta+\beta^{\ast}))+\frac{\kappa}{2}} , \beta=\frac{i g_{0} \alpha^{2}}{i \omega_{m}+\frac{\gamma}{2}} $). The algebraic equation for fixed point $ \alpha_s $ is a polynomial equation of degree 3, which has 2 stable and 1 unstable root \cite{Ghobadi32, Bahrampour42}. In order to obtain a higher coupling coefficient, the highest intensity with a stable fixed point is chosen. The input laser phase is opted such that $ \alpha_{s} $ is real.
The dynamics of the fluctuating operators around these steady states are as follows:
\begin{eqnarray}\label{Eq.6}
\frac{d \delta\hat{a}}{d t}&=& -i \Delta \delta\hat{a}-i \dot{\varphi}\alpha_{s}+i g_{0} \alpha (\delta \hat{b}+\delta \hat{b}^{\dagger})
\nonumber \\
&-&\frac{\kappa}{2}\delta\hat{a}+\sqrt{\kappa} \hat{\tilde{a}}_{in}
\nonumber \\
\frac{d  \delta\hat{b}}{d t}&=& -i \omega_{m} \delta\hat{b}+i g_{0} \alpha_{s} (\delta\hat{a}+\delta\hat{a}^{\dagger}) -\frac{\gamma}{2}\delta \hat{b}+\sqrt{\gamma} \hat{b}_{in}
\end{eqnarray}
where $\Delta=\Delta_{0}-g_{0} (\beta+\beta^{\ast})   $ is the effective cavity laser detuning. We work in the resolved sideband regime and have a good cavity ($ \kappa\ll \omega_{m} $). In the rotating wave approximation (RWA), by choosing $ \Delta =\omega_{m} $, the dynamical equations will be written as:
\begin{eqnarray}\label{Eq.7}
\frac{d \hat{a}}{d t}&=&-i \omega_{m} \hat{a} -i \dot{\varphi}\alpha_{s} +i G \hat{b} -\frac{\kappa}{2}\hat{a}+\sqrt{\kappa} \hat{a}_{in}
\nonumber \\
\frac{d \hat{b}}{d t}&=&-i \omega_{m} \hat{b}+ i G \hat{a} -\frac{\gamma}{2}\hat{b}+\sqrt{\gamma} \hat{b}_{in}
\end{eqnarray}
where $ G=g_{0}\alpha_{s} $ is the coupling coefficient between optical and mechanical field and $\hat{a},\hat{b} $ will be used as the linearised operators from now on.
We define quadratures of each mode as: $ \hat{x}_{i}=(\frac{A_{i}+A_{i}^{\dagger}}{2}) ,  \hat{p}_{i}=(\frac{A_{i}-A_{i}^{\dagger}}{2i})$ where $ i=1,2 $ and $ A_{1}= \hat{b} ,A_{2}= \hat{a}  $. Using Eq. (\ref{Eq.7}) one can obtain the evolution of the quadratures in the matrix form:
\begin{equation}\label{Eq.8}
\frac{d \hat{\vec{X}}}{d t}=Q\hat{\vec{X}}+\hat{\vec{X}}_{in}
\end{equation}
where $ \hat{\vec{X}}\in\Re^{4} $ is the vector of the quadratures :  $ \hat{\vec{X}} =[\hat{x}_{1},\hat{p}_{1},\hat{x}_{2},\hat{p}_{2}]^{T} $ and $\hat{\vec{X}}_{in} \in\Re^{4}$ is the input noise operators vector: $\hat{\vec{X}}_{in}=[\sqrt{\gamma}\hat{x}_{1,in},\sqrt{\gamma}\hat{p}_{1,in},\sqrt{\kappa}\hat{x}_{2,in},\sqrt{\kappa} \hat{p}_{2,in}-\alpha_{s} \dot{\varphi}]^{T} $. $ \hat{x}_{i,in}, \hat{p}_{i,in} (i=1,2)$ are defined by $  \hat{x}_{i,in}=(\frac{B_{i}+B_{i}^{\dagger}}{2}) ,
\hat{p}_{i,in}=(\frac{B_{i}-B_{i}^{\dagger}}{2i}) $ with $  B_{1}= \hat{b}_{in} ,B_{2}= \hat{a}_{in} $ and $ Q $ is given by:
\begin{eqnarray}\label{Eq.9}
Q = \left(\begin{array}{cccc}
-\dfrac{\gamma}{2} & \omega_{m} & 0 & -G \\
-\omega_{m} &  -\dfrac{\gamma}{2} & G & 0 \\
0 &  -G & -\dfrac{\kappa}{2} & \omega_{m} \\
G & 0 & -\omega_{m} & -\dfrac{\kappa}{2}
\end{array}  \right)
\end{eqnarray}

\subsection{The noise model}
The statistical properties of $ x_{i,in}, p_{i,in} (i=1,2)$ are defined by using Eq. (\ref{Eq.4}). The phase noise of the laser is responsible for the Laser linewidth $ \Gamma_L $. By taking the statistical properties of the laser phase noise as a white noise, the correlation function of the frequency noise will be $ \langle \dot{\varphi}(t) \dot{\varphi}(\acute{t})\rangle = 2 \Gamma_L \delta (t-\acute{t})  $ and a flat noise spectrum is obtained $ S_{\dot{\varphi}} (\omega)= 2 \Gamma_L $. However due to the finite non zero correlation time of frequency noise, the frequency noise spectral density can not be flat. In other words, it has a finite bandwidth color noise. In the simplest model, the spectrum is equivalent to a low pass filtered white noise with the following spectrum and correlation function \cite{Rabl31}:
\begin{equation}\label{Eq.10}
  S_{\dot{\varphi}} (\omega)= \frac{ 2 \Gamma_L}{1+\frac{\omega^{2}}{\gamma_{c}^{2}}} \ \ \ \  \langle \dot{\varphi}(t) \dot{\varphi}(\acute{t})\rangle = \Gamma_L \gamma_{c} e^{-\gamma_{c}\vert t-\acute{t} \vert}
\end{equation}
Here $ \gamma_{c} $ is the inverse of frequency noise correlation time or cutoff frequency in the laser frequency noise spectrum, so that the frequency noise is suppressed at frequencies $ \omega >\gamma_{c}  $. By increasing $ \gamma_{c} $, the correlation time decreases and the frequency noise starts reaching the white noise. This frequency spectrum corresponds to the differential equation $ \ddot{\varphi}(t)+\gamma_{c}\dot{\varphi}(t)=\varepsilon(t) $ where $ \varepsilon(t) $ is a Gaussian random variable with white noise correlation function \cite{Ghobadi32}:
\begin{equation}\label{Eq.11}
  \langle \varepsilon(t) \varepsilon(\acute{t})\rangle = 2\gamma_{c}^{2}\Gamma_{L} \delta(t-\acute{t})
\end{equation}
In the following calculations, it is more convenient to use input white noise rather than color noise, so we define an auxiliary operator $ \psi\equiv\dot{\varphi} $ and add its corresponding differential equation ($  \dot{\psi}(t)+\gamma_{c}\psi(t)=\varepsilon(t)  $) to our four coupled differential equations of quadratures. The quadratures vectors $  \hat{\vec{X}}  $ and $  \hat{\vec{X}}_{in}  $ are extended to $   \hat{\vec{X}}_{E}=[\hat{x}_{1},\hat{p}_{1},\hat{x}_{2},\hat{p}_{2}, \psi]^{T} \in \Re^{5}$ and $  \hat{\vec{X}}_{in,E}=[\sqrt{\gamma}\hat{x}_{1,in},\sqrt{\gamma}\hat{p}_{1,in},\sqrt{\kappa}\hat{x}_{2,in},\sqrt{\kappa} \hat{p}_{2,in}, \varepsilon]^{T}  \in \Re^{5} $, respectively and Eq. (\ref{Eq.8}) is changed to $ \frac{d \hat{\vec{X}}_{E}}{d t}=Q_{E}\hat{\vec{X}}_{E}+\hat{\vec{X}}_{in,E} $ with:
\begin{eqnarray}\label{Eq.12}
Q_E = \left(\begin{array}{ccccc}
-\frac{\gamma}{2} & \omega_{m} & 0 & -G & 0\\
-\omega_{m} &  -\frac{\gamma}{2} & G & 0 & 0\\
0 &  -G & -\frac{\kappa}{2} & \omega_{m} & 0\\
G & 0 & -\omega_{m} & -\frac{\kappa}{2} & -\alpha_{s} \\
 0 &  0 &  0 &  0 &  -\gamma_{c}
\end{array}  \right)
\end{eqnarray}
In the following section, the method to finding the fidelity of the final state with the initial state, in the presence of coupling laser is explained.

\section{Gaussian state fidelity}
For the sake of simplicity, it is assumed that the initial state of the cavity is a pure Gaussian state. At this step we are going to find the fidelity of the optical state at time $ t $ with a given initial state, while the optomechanical interaction is switched on. Our approach is the same as the one used in reference \cite{Wang25}.
The fidelity between initial and final state is given by $ F=Tr(\hat{\rho}_{i}\hat{\rho}_{f}) $, where $ \rho_{i} $ and $ \rho_{f} $ are the density matrices of the initial and final states, respectively. In Wigner representation, the fidelity can be written as $ F=\pi \int _{-\infty}^{\infty} d \vec{\xi} W_{i}(\vec{\xi}) W_{f}(\vec{\xi}) $,
where $ \vec{\xi} \in\Re^2$ is the vector of the optical quadratures $ \vec{\xi}=[\hat{x_{2}},\hat{p_{2}}]^{T} $. The Wigner function of Gaussian states can be written in terms of covariance matrix of quadratures $ V $, where the symmetrized covariance matrix is defined by $ V_{ij} =\frac{1}{2}\langle \hat{ X_{i}} \hat{ X_{j}}+\hat{ X_{j}} \hat{ X_{i}}\rangle$. Due to the bilinear form of interaction  Hamiltonian, the final state remains Gaussian. So after some mathematical manipulation, a relation for Gaussian state fidelity is achieved:
\begin{equation}\label{Eq.13}
F=\frac{1}{2\sqrt{V_{i}+V_{f}}}e^{\lbrace -\frac{1}{2}(\langle\hat{\vec{\xi}}_{i}\rangle-\langle\hat{\vec{\xi}}_{f}\rangle)\cdot \frac{1}{V_{i}+V_{f}}(\langle\hat{\vec{\xi}}_{i}\rangle-\langle\hat{\vec{\xi}}_{f}\rangle) \rbrace}
\end{equation}
Indices $ i $ and $ f $ stands for initial and final states, respectively. It is noted that the fidelity must be maximized by employing an orthogonal transformation in phase space, so if $ \rho_{f} $ is the rotated version of $ \rho_{i}  $, the storing fidelity will be 1.

Two useful parameters $ \bar{n}_{h} =2 \sqrt{det(V_{i}+V_{f})}-1 $ and  $ \lambda^{2}=(\langle\hat{\vec{\xi}}_{i}\rangle-\langle\hat{\vec{\xi}}_{f}\rangle) \cdot  \frac{\sqrt{det(V_{i}+V_{f})}}{V_{i}+V_{f}}(\langle\hat{\vec{\xi}}_{i}\rangle-\langle\hat{\vec{\xi}}_{f}\rangle)  $ is introduced to make the fidelity formula more straightforward \cite{Wang25}:
\begin{equation}\label{Eq.14}
F=\frac{1}{1+\bar{n}_{h}}e^{\frac{-\lambda^{2}}{1+\bar{n}_{h}}}
\end{equation}
The parameter $ \bar{n}_{h} $ which is the effective number of thermal quantas, originates from input noises and heats the state throughout the process. The parameter $ \lambda $ shows the effect of optical and mechanical damping rates on the fidelity of the state. In order to study the fidelity of the retrieved state, the expectation value of optical quadratures and also covariance matrix at initial and final states, must be calculated.
The initial covariance matrix is obtained easily by knowing the initial state of the optical and mechanical mode. The evolution of the covariance matrix can be derived from Eq. (\ref{Eq.8}) (see Appendix A):
\begin{equation}\label{Eq.15}
\frac{d V}{dt}=Q_{E}V+VQ_{E}^{T}+N
\end{equation}
where $ N $ is the diffusion matrix and can be achieved for color noise as (See Appendix A):
\begin{eqnarray}\label{Eq.16}
 N = \left(\begin{array}{ccccc}
\frac{\bar{\gamma}}{4} & 0 & 0 & 0 & 0\\
0 & \frac{\bar{\gamma}}{4} & 0& 0 & 0\\
0 &  0& \frac{\bar{\kappa}}{4} & 0 & 0\\
0 & 0 & 0 &  \frac{\bar{\kappa}}{4}& 0 \\
0 & 0 & 0 & 0 & 2\gamma_{c} ^{2}\Gamma_{L}
\end{array}  \right)
\end{eqnarray}
 $ \bar{\gamma} $ and $ \bar{\kappa} $ are defined by $ \bar{\gamma}=\gamma (1+2 N_{m}) $ and $ \bar{\kappa}=\kappa (1+2 N_{C}) $, respectively. Eq. (\ref{Eq.15}) is an inhomogeneous differential matrix equation that can be written as a system of 25 coupled ordinary differential equations. Due to the symmetry in the covariance matrix, only 15 of these equations are independent. By solving these equations by ordinary methods like Runge–Kutta, the covariance matrix of the final state can be derived. However, the analytic solution of this equation is as follows:
\begin{equation}\label{Eq.17}
V(t)=e^{Q_{E}t} V(0) e^{Q_{E}^{T}t}+\int_{0}^{t} d\acute{t} e^{Q_{E}(t-\acute{t})} N  e^{Q_{E}^{T}(t-\acute{t})}
\end{equation}
It is assumed that the average value of input noises are zero. Averaging over Eq. (\ref{Eq.8}), the input noises are eliminated and the equation is reduced to : $ \frac{d \langle \hat{\vec{X}} (t)\rangle }{dt}= Q \langle \hat{\vec{X}} (t)\rangle $, which has the following solution:
\begin{equation}\label{Eq.18}
\langle \hat{\vec{X}} (t)\rangle = e^{Qt}\langle \hat{\vec{X}} (0)\rangle
\end{equation}
Note that due to elimination of noise, we don't have to use the extended version of $ \hat{\vec{X}} $ and $ Q $. Using Eqs. (\ref{Eq.17}),(\ref{Eq.18}) and (\ref{Eq.13}) , the optical state fidelity at time $ t $ with initial state can be derived. As it has been mentioned before, the terminal fidelity should be maximized. Using the unitary transformation $ e^{-Q_{r}t} $ to rotate the parameters on the phase space, the fidelity is maximized, where $ Q_{r} $ is given by:
\begin{eqnarray}\label{Eq.19}
 Q_{r}= \left(\begin{array}{ccccc}
0& \omega_{m} & 0 & 0 & 0\\
-\omega_{m} & 0 & 0& 0 & 0\\
0 &  0& 0 & \omega_{m} & 0\\
0 & 0 & -\omega_{m} & 0& 0 \\
0 & 0 & 0 & 0 & 0
\end{array}  \right)
\end{eqnarray}
If the initial optical state is not a Gaussian state, such as fock state and any finite linear combination of fock states, it would be difficult to find analytic expression for the  fidelity. Generally the initial density matrix of the system can be written as  $ \rho(0)=\rho_{o}\otimes \rho_{m} $ , where indices $ o $ and $ m $ indicate optical and mechanical modes, respectively. The evolution of density matrix of the system can be obtained by the numerical solution of the Lindblad master equation in the presence of optical and mechanical dissipations and input noises \cite{Wiseman43, Joshi44}. The final density matrix can be used in $  F=Tr(\hat{\rho}(0)\hat{\rho}(t)) $ to give the fidelity between initial and final states.

\subsection{Storing Protocol}
The storing fidelity depends on the input state. Memory fidelity can be defined as the average of fidelities over the possible input states. In this paper our study is concentrated on the squeezed coherent state: $ \ket{\alpha,r}=D(\alpha) S(r) \ket{0}$ where $ D(\alpha)=e^{\alpha \hat{a}^{\dagger}-\alpha^{\ast}\hat{a}} $ and $ S(r)=e^{\frac{r}{2}(\hat{a}^{2}-\hat{a}^{\dagger 2})} $ are displacement and squeezing operators, respectively. It is assumed that $ r $ is the real squeezing parameter. By setting $ r=0 $, we will have a coherent state. The initial state of the oscillator is assumed to be a ground state $ \ket{0} $, which means that the oscillator has been cooled. This assumption is almost possible by using a different cooling method \cite{Gigan12, Wilson-Rae13, Schliesser14, Chan15} or using an optomechanical coupling $ \frac{\pi}{2} $ pulse before the storing protocol, which is actually cooling the oscillator near the quantum ground state \cite{Tian23}. If the cooling process hasn't been done, the initial state of the mechanical mode would be the thermal state. The storing protocol is composed of three steps:

1) At initial time, the average of the quadratures are $ \langle \vec{X(0)}\rangle=[0,0,Re(\alpha),Im(\alpha)]^{T} $. The initial covariance matrix is:
\begin{equation}\label{Eq.20}
V(0)= \frac{1}{4} \ \left(\begin{array}{ccccc}
1& 0 & 0 & 0 & 0\\
0 &  1 & 0 & 0 & 0\\
0 &  0 & e^{-2r} & 0 & 0\\
0 & 0 & 0 & e^{2r} & 0 \\
0 & 0 & 0 & 0 & 1
\end{array}  \right)
\end{equation}
If the initial state of the oscillator hasn't been cooled, two elements of the covariance matrix $ V $ ($ V_{11} $ and $ V_{22} $), are changed to $ 1+2N_{m} $, where $ N_{m} $ is the mean thermal phonon number as defined before. By switching on the coupling laser for the duration time $ t_{s}=\frac{\pi}{2G} $ (write pulse),  the state of the optical mode is almost swapped with the mechanical mode. The quadratures average and covariance matrix can be obtained  from Eqs. (\ref{Eq.17}) and (\ref{Eq.18}).

2) The coupling laser is switched off and the optical state is stored in the mechanical mode for a time $ \tau $. In this step, the $ Q $ matrix is changed to $ Q^{\prime} $, where it is the same as matrix  $ Q $ except that the elements containing $ G $, go to zero.

3) After the storage time is up, by switching on the coupling $ \frac{\pi}{2} $  pulse again (read pulse), the state that is stored in the mechanical mode, is mapped to the optical mode and can be retrieved from the cavity. The second $ \frac{\pi}{2} $  pulse must be $ \pi $ radians out of phase relative to the first one to cancel the accumulated phase during the process \cite{McGee29}. The schematic figure of cooling, writing and reading pulses are shown in Fig. 2.

\begin{figure}[htbp]
\centerline{\includegraphics[width=1\columnwidth]{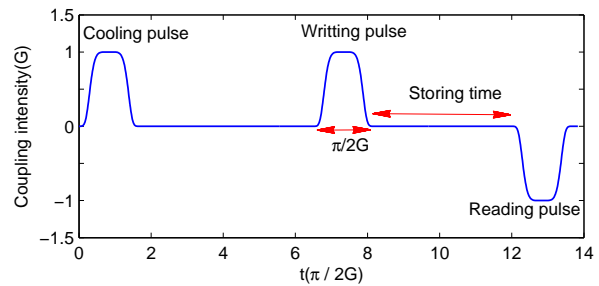}}
  \caption{(color online)
  A schematic outline of coupling $ \frac{\pi}{2} $ pulse for quantum memory. The first pulse applies for cooling the mirror near ground state. This pulse will change the state of initial optical mode (vacuum state) with thermal mechanical mode. After a while, the intracavity thermal optical mode transfers to the reservoir and the input Gaussioan state enters the cavity at the beginning of the write pulse. After the storage time $ \tau $, the reading $  \frac{\pi}{2} $ pulse with an opposite sign of the writing pulse, enters the cavity.}
\end{figure}

The final quadratures average and covariance matrix is obtained from:

\begin{equation}\label{Eq.21}
\langle \hat{\vec{X}}_{f} \rangle =  e^{Q^{\prime \prime}t_{s}}e^{Q^{\prime}\tau}e^{Qt_{s}}\langle \hat{\vec{X}}_{i}\rangle
\end{equation}
and

\begin{eqnarray}\label{Eq.22}
V_{f}&=&e^{Q^{\prime \prime}_{E}t_{s}} e^{Q^{\prime}_{E}\tau} e^{Q_{E}t_{s}}V_{i} e^{Q_{E}^{T}t_{s}}e^{Q^{\prime T}_{E}\tau} e^{Q^{\prime \prime T}_{E}t_{s}}
\nonumber \\
&+&\int_{0}^{t_{s}} d\acute{t} e^{Q^{\prime \prime}_{E}t_{s}} e^{Q^{\prime}_{E}\tau} e^{Q_{E}(t_{s}-\acute{t})}N e^{Q_{E}^{T}(t_{s}-\acute{t})}e^{Q^{\prime T}_{E}\tau} e^{Q^{\prime \prime T}_{E}t_{s}}
\nonumber \\
&+&\int_{0}^{\tau} d\acute{t}e^{Q^{\prime \prime}_{E}t_{s}} e^{Q^{\prime}_{E}(\tau-\acute{t})} N e^{Q^{\prime T}_{E}(\tau-\acute{t})}e^{Q^{\prime \prime T}_{E}t_{s}}
\nonumber \\
&+&\int_{0}^{t_s} d\acute{t}e^{Q^{\prime \prime}_{E}(t_{s}-\acute{t})} N e^{Q^{\prime \prime T}_{E}(t_{s}-\acute{t})}
\end{eqnarray}
where matrix $ Q^{\prime \prime} $ and $ Q^{\prime \prime}_{E} $ is obtained by replacing $ G $ by $ -G $ in matrix $ Q $ and $  Q_{E} $, respectively. (Due to the phase difference between read and write pulses). The fidelity is obtained by Eq. (\ref{Eq.13}). Effects of the environment are considered in cavity and mechanical decay rates and also in thermal phonon $ N_m $ and photon $ N_C $ numbers. Effects of control laser noise on the memory fidelity are given by presence of laser bandwidth $ \Gamma_{L} $ and frequency spectral density bandwidth $ \gamma_{c} $.

\section{Numerical results}
To investigate the effect of coupling (control) signal noise on the memory fidelity, numerical calculations are done for a microwave cavity and an optomechanical cavity in the following subsections.
\subsection{microwave cavity}

To consider the effect of coupling signal phase noise on the storing fidelity in an electromechanical system, the parameters are chosen as those employed in references \cite{Teufel39, McGee29}. These parameters correspond to a microwave cavity. In the next subsection the optomechanical Fabry-Perot cavity, is considered. The resonance frequency of the mechanical mode is $ \frac{\omega_{m}}{2\pi} =10.69\ MHz$ and the cavity decay rate is chosen to be $ \frac{\kappa}{2\pi} =170\ kHz $ \cite{Teufel39}. We are working in the strong coherent coupling regime of operation ($ \gamma_{th} \ll \kappa\ll G $) and so we choose the largest value of electromechanical coupling constant $ G $ that has been achieved in experiments to date ($ G=0.05\times \omega_{m}=2\pi \times 534.5\ kHz $) \cite{McGee29}. The initial value of electromagnetic mode that we are interested in storing, is the coherent state $ \ket{\alpha} $ with $ \alpha=1 $. The single photon coupling strength $ g_0 $ is $ {2\pi} \times 230\ Hz $ which is obtained by optimizing both $ \frac{\partial \omega_C}{\partial L} $ and $ x_{zpf} $ to achieve large values of $ g_0 $ \cite{Teufel39}. The mean thermal phonon number is taken to be $ N_m=3 $, which corresponds to the temperature $ 1.7\ mK $ and the storage time is set to $ \tau=64\ {\omega_{m}}^{-1}=0.95\ \mu s $. By changing the mechanical quality factor of the system $ Q_{m}=\frac{\omega_{m}}{\gamma}$ ($  \omega_{m} $ is fixed while $ \gamma $ is changed), the storing fidelity $ F $, $ n_h $ and $ \lambda $, are calculated and plotted for different values of microwave control linewidth $ \Gamma_L $ and cutoff frequencies $ \gamma_{c} $ in Figs. 3(a), 3(b) and 3(c), respectively.
By increasing the noise intensity in the electromechanical system, the parameters $ \bar{n}_h $ and $ \lambda $ are increasing which leads to the drop in fidelity.
\begin{figure}[htbp]\label{fig.3}
\centerline{\includegraphics[width=1\columnwidth]{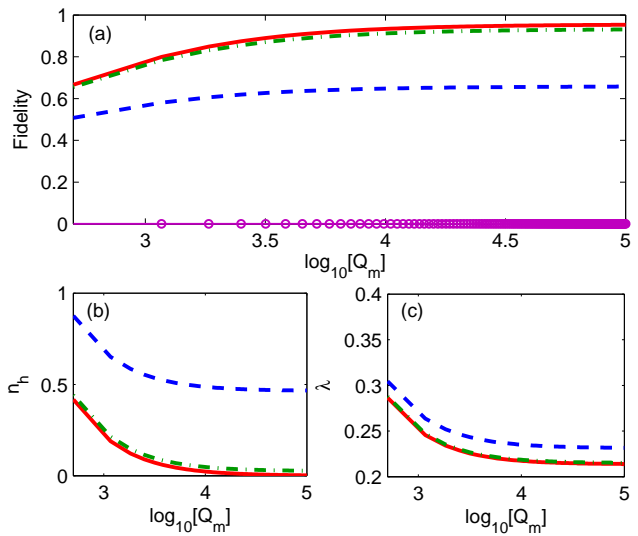}}
  \caption{(color online)
Variation of a) Storing fidelity, b) heating parameter $ \bar{n}_h $ ans c) damping parameter $ \lambda $, versus mechanical quality factor $Q_{m}=\frac{\omega_{m}}{\gamma}$ ($ \omega_{m}=2\pi \times 10.69\ MHz $), for $ \Gamma_L=0  $ (solid red line), $ \Gamma_L=1\ kHz, \gamma_c=100 \ kHz  $ (dash-dot green line), $ \Gamma_L=1\ kHz, \gamma_c=0.5 \ MHz  $ (dashed blue line) and $ \Gamma_L=1\ kHz,\gamma_c=10 \ MHz$ ( $  \gamma_c\rightarrow \infty $ or white noise) (purple line with marker  $ \circ $ ) (white noise hasn't been shown in (b,c)).
The input state is a coherent state $ \ket{\alpha} $ with $ \alpha=1 $, the thermal phonon number is $ N_{m}=3 $, and the storage time is set to $ \tau=\frac{64}{\omega_{m}} $.}
\end{figure}
The amplitude and bandwidth of phase noise spectral density are proportional to the control linewidth $ \Gamma_{L} $ and cutoff frequency $ \gamma_{c} $, respectively. Hence, the fidelity decreases by increasing $  \Gamma_{L} $ and $ \gamma_{c}$. As it is expected, Fig. 3(a) shows that, by increasing the cutoff frequency, the fidelity will mainly degrade. However, for the small values of the phase noise parameters like $ \Gamma_L=1\ kHz $ and $\gamma_c=100 \ kHz $ (dash-dot green line in Fig. 3(a)), the phase noise will not affect the fidelity significantly. The limit $ \gamma_{c}\rightarrow \infty $ corresponds to the white noise, which shows very low fidelity nearly zero (purple line with marker $ \circ $). Fidelity is an increasing function of mechanical quality factor $ Q_{m} $. Due to decoupling of mechanical oscilattor from its environment in high values of $ Q_{m} $, the effective number of thermal quantas that heat the state decreases, which results in decreasing the parameter $ \bar{n}_h $ (see Fig. 3(b)). Furthermore, by decreasing mechanical decay rate $ \gamma $, the damping parameter $ \lambda $ decreases (see Fig. 3(c)), and as a result, the fidelity increases.

For $ Q_m> 10^4 $, the storing fidelity for low values of noise parameters (dash-dot green line in Fig. (3.a)) is more than $ F>0.8 $, which is suitable for experimental investigations. However, for low values of $ Q_m $, where the mechanical decay rate becomes large, the final state will thermalize faster and the fidelity declines.

For a given mechanical quality factor of the oscillator $ Q_m=360000 $ (which corresponds to $ \gamma=186.57\ s^{-1} $), the effect of the signal control linewidth $ \Gamma_L $ and noise bandwidth $ \gamma_{c} $ in the fidelity is calculated and presented in Fig. (4.a) and Fig. (4.b), respectively. As it is expected, the fidelity is a decreasing function of signal linewidth $ \Gamma_L $, while in terms of cutoff frequency $ \gamma_{c} $, it decreases and approaches saturation corresponding to the white noise value. Variation of fidelity as a function of cutoff frequency $ \gamma_{c} $, for three values of $ \Gamma_L=1\ kHz, 5\ kHz, 10\ kHz $ , is presented in Fig. (4.b).
\begin{figure}[htbp]\label{fig.4}
\centerline{\includegraphics[width=1\columnwidth]{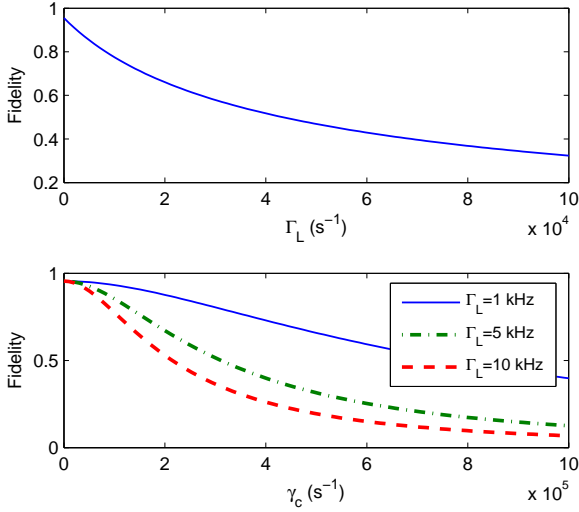}}
  \caption{(color online)
a) Variation of fidelity as a function of control linewidth $ \Gamma_L $. b) The variation of fidelity as a function of the control cutoff frequency $ \gamma_{c} $ for $ \Gamma_L=1\ kHz $ (solid blue line),  $ \Gamma_L=5\ kHz $ (dash-dot green line) and $ \Gamma_L=10\ kHz $ (dashed red line). The input state is a coherent state $ \ket{\alpha} $ with $ \alpha=1 $. $ Q_m=360000 $.}
\end{figure}

The control phase noise is present only during the write and read process, while the environmental noises of optomechanical system exist in all the read, write and storage times. Fig. 5 shows the storing fidelity as a function of the storage time $ \tau $, for two values of reservoir temperature in the presence and in the absence of control phase noise effect. As the storage time increases, the duration time that mechanical mode interacts with thermal noise increases, which results in reducing fidelity.
\begin{figure}[hbp]\label{fig.5}
\centerline{\includegraphics[width=0.95\columnwidth]{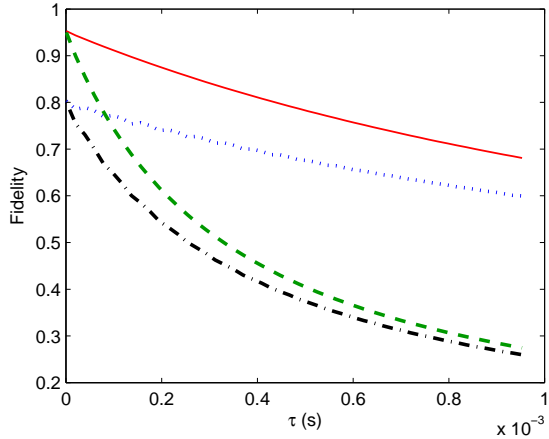}}
  \caption{(color online)
Storing fidelity as a function of the storing time $ \tau $  for different values of reservoir temperature, laser linewidth $  \Gamma_L $ and cutoff frequency $ \gamma_c $ : $ T=1.7\ mK, \Gamma_L=0  $ (solid red line), $T=1.7\ mK, \Gamma_L=1\ kHz, \gamma_c=300\ kHz  $ (doted blue line), $T=0.01\ mK, \Gamma_L=0 $ (dashed green line) and $ T=0.01\ mK, \Gamma_L=1\ kHz, \gamma_c=300\ kHz $  (dash-dot black line). The input state is a coherent state $ \ket{\alpha} $ with $ \alpha=1 $.}
\end{figure}
Reservoir temperature plays an important role in the storing fidelity. If the reservoir temperature is $ 1.7 mK $, the memory will store the initial coherent state up to $ 0.4 \ ms $ with fidelity more than $ 0.8 $ (in the absence of control noise effect) and with fidelity more than $ 0.7 $ (in the presence of control noise effect). By increasing the reservoir temperature to $ 0.01\ K$, the storage time drops to $ 0.05\ ms $ for an appropriate fidelity.
As it is expected, Fig. 5 shows that by increasing the storage time, the effect of control phase noise on fidelity decreases relative to other environmental noises and the curves considering the control phase noise effect, reach the corresponding curve without this effect. That is because the control phase noise influences memory only during the read and write pulses and doesn't have any effect on the fidelity in the storage time.

To consider the effect of coupling coefficient $ G=\alpha_{s}g_0 $ on the fidelity of memory, an electromechanical system with the following parameters is taken into consideration. The mechanical quality factor of the oscillator is fixed and is equal to $ Q_{m}=360000 $. By changing the coupling coefficient $ G $ from $ 0.02 \times \omega_{m} $ to $ 0.05 \times \omega_{m} $, the fidelity of storing the coherent state $ \ket{\alpha} $ for storage time $ \tau=\frac{64}{\omega_{m}} $, is calculated and represented in Fig. 6(a). The coupling coefficient $ G $ is adjusted by $ \alpha_{s} $. So, Fig. 6(a) also shows variation of memory fidelity versus $ \alpha_{s}  $.
\begin{figure}[htbp]\label{fig.6}
\centerline{\includegraphics[width=1\columnwidth]{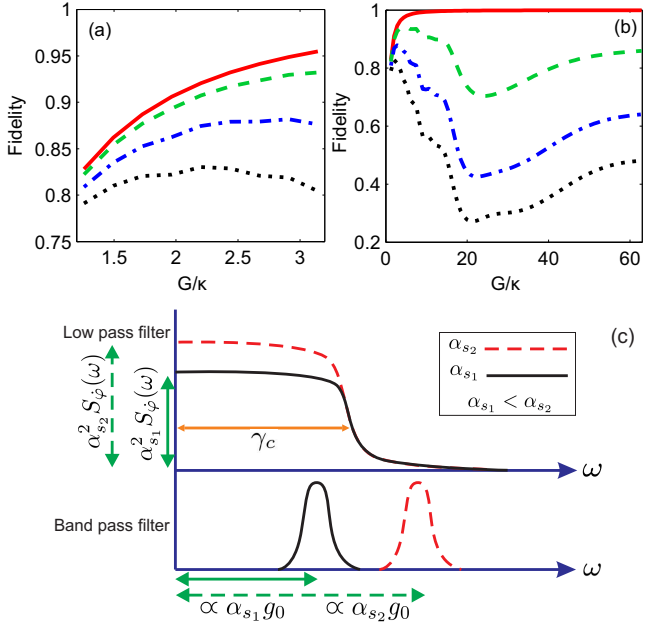}}
  \caption{(color online)
a, b) Storing fidelity as a function of $ \frac{G}{\kappa} $ for two ranges of variation of $ G $ (a,b) , in the absence and in the presence of laser phase noise effect: $ \Gamma=0  $ (solid red line), $ \Gamma=1\ kHz, \gamma_c=100\ kHz  $ (dashed green line), $ \Gamma=1\ kHz, \gamma_c=200\ kHz  $ (dash-dot blue line) and $ \Gamma=1\ kHz, \gamma_c=300\ kHz  $  (dotted black line). c) Schematic description of cascading a low pass filter with cutoff frequency $ \gamma_{c} $ and a band pass filter with center frequency determined by the coupling coefficient $ G $ for two values of coupling coefficient $ G_1 \varpropto \alpha_{s 1} $ and $ G_2 \varpropto \alpha_{s 2} $. $ \alpha_{s 2}  >\alpha_{s 1} $. The storage time is $ \tau=\frac{64}{\omega_{m}} $.}
\end{figure}
As already expected, in the absence of control phase noise (solid red line in Figs. 6(a) and 6(b)), the storing fidelity increases by increasing the coupling coefficient between microwave and mechanical modes. In this situation, the interaction time $ t_s $ decreases as $ G $ increases, which results in less thermal noise affecting the system in the interaction time. However in the presence of control phase noise, the damaging effect of phase noise is more effective in higher coupling coefficient $ G $, which results in the decrease of the storing fidelity in high values of $ G $. This phenomenon originates from the coefficient $ \alpha_{s} $ of the phase noise term in Eq. (\ref{Eq.7}). So, by increasing the input coupling (control) intensity, despite of the drop in interaction time $ t_s $, the effect of input control phase noise on the system gets stronger, which prevents the fidelity from increasing. According to the mean value theorem, there is a maximum in the fidelity versus $ \alpha_{s} $ in the presence of control phase noise, which can be observed obviously in Fig. 6(a).

Let us now increase the coupling coefficient $ G $ more than has been achieved in experiments. Fig. 6(b) shows the variation of memory fidelity versus coupling coefficient, where $ G $ is changed from $ 0.02 \times \omega_{m} $ to $  \omega_{m} $.
It is easy to show that in the strong coupling regime of operation, an optomechanical system is equivalent to a tune filter for the control phase noise, and its center frequency can be determined by the coupling coefficient $ G $. Hence the control phase noise can be modelled by cascading a low pass filter with cutoff frequency $ \gamma_{c} $ and a band pass filter with a central frequency determined by the coupling coefficient $ G $ (see Fig. 6(c)). Increasing  $ \alpha_{s} $ increases the noise intensity and consequently decreases the memory fidelity. By  increasing $ \alpha_{s} $ even more, the tune filter central frequency moves away from the cutoff frequency of low pass filter and the noise intensity decreases due to normal mode splitting. In other words, the read and write duration time decreases below the coherence time of control frequency noise and as a result the memory fidelity increases. The minimum observed in Fig. 6(b) is related to this effect. For large $ \alpha_{s} $, the effect of control phase noise is negligible and fidelity is affected by the optomechanical system's environmental interactions.

In order to compare the storing fidelity of the squeezed state with the coherent state, the squeezed state fidelity $ \ket{\alpha,r} $ for four values of $ r=0, 0.2,0.5,0.8 $, as a function of $ Q_m $ is calculated and shown in Fig. 7(a). 
\begin{figure}[htp]\label{fig.7}
\centerline{\includegraphics[width=1\columnwidth]{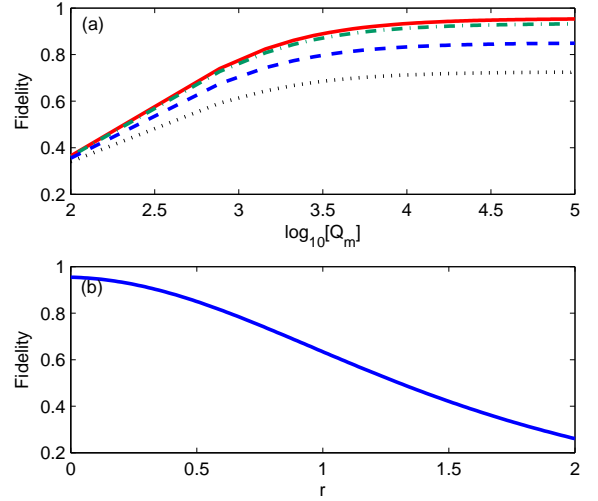}}
  \caption{(color online)
a) Storing fidelity as a function of mechanical quality factor $ Q_m $ with $ \Gamma_L=1\ kHz, \gamma_c=10\ kHz  $ for different values of squeezing  parameter $ r $:  $ r=0  $ (solid red line), $r=0.2 $ (dash-dot green line), $ r=0.5  $ (dashed blue line) and $ r=0.8  $ (dotted black line). b) Variation of fidelity as a function of squeezing parameter $ r $ for fixed vaue of  $ Q_m=360000 $. The storage time is set to $ \tau=0.95\ \mu s $.}
\end{figure}
The storage time is set to $ \tau=0.95\ \mu s $ and other parameters are chosen like those are used in Fig. 3. The values of $ \Gamma_L, \gamma_{c} $ is taken $  \Gamma_L=1\ kHz $ and $ \gamma_{c}=10\ kHz $, respectively. Fig. 7(b), shows the variation of fidelity as a function of squeezing parameter $ r $, when mechanical quality factor is fixed to $ Q_m=360000 $.
It is clear that the fidelity decreases as parameter $ r $ increases while the initial state becomes more and more non classical. By growing the squeezing parameter $ r $, the initial state would be more sensitive to input thermal and optical noises \cite{Wang25}.

\subsection{Optomechanical cavity}
The optomechanical cavity that we consider is a Fabry-Perot cavity with strong coupling between optical and mechanical mode \cite{Groblacher40}. The cavity and mechanical decay rates are $ \frac{\kappa}{2\pi} = 215\ kHz $ and $ \frac{\gamma}{2\pi} = 140\ Hz  $, respectively. The mechanical frequency of micromirror is $ \frac{\omega_{m}}{2\pi}=947\ kHz $ and the single photon coupling strength is $ g_0=2\pi \times 1.91\ Hz $. The input coupling Nd:YAG laser with power $ 11\ mW $ and wavelength $ 1064\ nm $ enters the cavity and makes optomechanical coupling coefficient $ G ={2\pi} \times 229.81\ kHz$. In the absence of laser phase noise, the fidelity of storing the coherent state $ \ket{\alpha} $ with $ \alpha=1 $, for storage time $ 0.95\ \mu s $, with thermal phonon number $ N_m=3 $, will be obtained $ 0.789 $.
By considering the effect of phase noise as a laser linewidth $ \Gamma_{L}=1\ kHz $ and cutoff frequency $ \gamma_{c}=1\  kHz $, the storing fidelity decreases to $ 0.28 $ which is an undesirable fidelity. In order to compare the fidelity of microwave and optomechanical systems, the storing fidelity for some values of noise parameters as a function of the mechanical quality factor (by fixing $ \omega_{m} $ and changing $ \gamma $) is presented in Fig. 8. The storage time and thermal phonon number have been chosen to be similar to Fig. 3  ($ \tau=0.95\ \mu s $ , $ N_m=3 $ ).
\begin{figure}[htbp]\label{fig.8}
\centerline{\includegraphics[width=0.85\columnwidth]{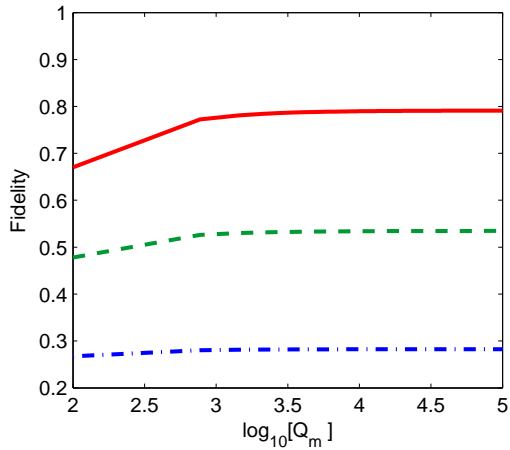}}
  \caption{(color online)
Storing fidelity as a function of mechanical quality factor $ Q_m $,  for $ \Gamma_L=0  $ (solid red line), $ \Gamma_L=1\ kHz, \gamma_c=0.5 \ kHz  $ (dashed green line), $ \Gamma_L=1\ kHz, \gamma_c=1 \ kHz  $ (dash-dot blue line).  Storage time and thermal phonon number are $ \tau=0.95\ \mu s $ and $ N_m=3 $, respectively.}
\end{figure}

Comparing Fig. 3 with Fig. 8, we find out that the degrading effect of phase noise on the fidelity of optomechanical system is more significant than that in the microwave system. This could be explained by considering the fact that, the single photon coupling strength $ g_0 $ of our microwave system ($ 2\pi \times 230\ Hz $) is more than that one in optomechanical system ($  {2\pi} \times 1.91\ Hz  $). So in order to achieve the same value of coupling coefficient $ G $, the lower intracavity amplitude $ \alpha_{s} $ is needed in microwave system. The reduction in $ \alpha_{s} $ leads to the reduction of phase noise effect in microwave system. On the other hand, the coupling coefficient $ G $
in microwave system ($ G=2\pi \times 534.5\ kHz $) is more than that in optomechanical system ($ {2\pi} \times 229.81\ kHz $) which gives the shorter interaction time between light and matter in microwave system and hence the fewer amount of thermal and phase noise, will affect the system.

\section{Conclusion}
The effect of laser phase noise on the fidelity of electromechanical and optomechanical quantum memories have been studied.
The phase noise of the laser behaves like a color noise which is included in the dynamics of the equations by means of an axillary variable with input white noise. It is shown that the damaging effect of laser phase noise on memory fidelity is proportional to the intracavity field amplitude $ \alpha_{s} $.
For the electromechanical system that we have considered \cite{Teufel39} in the strong coupling regime of operation, the storing fidelity of the coherent state $ \ket{\alpha} $ for an interval $ \tau\sim 1 \mu s $, drops from $ 0.95 $, in the absence of control phase noise effect, to $  0.66 $, in the presence of control phase noise, with a linewidth of $ \Gamma_{L}=1\ kHz $ and a cutuff frequency of $ \gamma_{c}=0.5\ MHz $. By decreasing the cut off frequency to $ \gamma_{c}=10\ kHz $, the memory fidelity increases significantly and the degrading effect of noise disappears.
It has been shown that by minimizing the noise effect, the coherent state can be stored up to $ 0.4\ ms $ with a fidelity more than $ 0.8 $, in mechanical mode. While by considering the effect of phase noise as a control line width $ 1\ kHz $ and cutoff frequency $ \gamma_{c}=300\ kHz $, the storage time with appropriate fidelity decreases to $ 0.05\ ms $.
By increasing the input control intensity, the optomechanical coupling $ G $ increases and the interaction time between photons and phonons decreases, which results in memory fidelity gain, while at the same time due to the increase in intracavity field amplitude, the effect of phase noise elevates which prevents the fidelity from increasing. By comparing the electromechanical and optomechanical systems in the range of experimentally achievable parameters \cite{Teufel39, Groblacher40}, it was found that, the system with higher single photon coupling $ g_0 $, is more robust against phase noise degradation effects.

\acknowledgements
We would like to thank Mrs. N. Pishbin.

\appendix
\section{}
The dynamic equations of quadratures vector are obtained from the extended version of Eq. (\ref{Eq.8}). Here we define variable: $ V_{1}=\langle \hat{\vec{X} } \hat{\vec{X}}^{T}\rangle$. It is easy to show that the covariance matrix is related to $ V_{1} $ through equation $ V= \frac{V_{1}+V_{1}^{T}}{2} $. Using Eq. (\ref{Eq.8}), the following equations are derived:
\begin{eqnarray}\label{App1}
\dot{\hat{\vec{X}}}  \hat{\vec{X}}^{T}&=&Q_{E} \hat{\vec{X} } \hat{\vec{X}}^{T}+\hat{\vec{X}}_{in} \hat{\vec{X}}^{T}
\nonumber \\
\hat{\vec{X}} \dot{\hat{\vec{X}}}^{T}&=&\hat{\vec{X} } \hat{\vec{X}}^{T} Q_{E}^{T}+\hat{\vec{X}}\hat{\vec{X}}_{in}^{T}
\end{eqnarray}
which gives us the relations:
\begin{eqnarray}\label{App2}
\frac{d V_{1}}{dt}&=&Q_{E}V_{1}+V_{1}Q_{E}^{T}+\hat{\vec{X}}_{in}\hat{\vec{X}}^{T}+\hat{\vec{X}}\hat{\vec{X}}_{in}^{T}
\nonumber \\
\frac{d V_{1}^{T}}{dt}&=&V_{1}^{T}Q^{T}+Q V_{1}^{T}+(\hat{\vec{X}}_{in}\hat{\vec{X}}^{T}+\hat{\vec{X}}\hat{\vec{X}}_{in}^{T})^{T}
\end{eqnarray}

By adding these two equations, the dynamic equation of the covariance matrix is achieved:
\begin{equation}\label{App3}
\frac{dV}{dt}=QV+VQ^{T}+\frac{\langle\hat{\vec{X}}_{in}\hat{\vec{X}}^{T}+\hat{\vec{X}}\hat{\vec{X}}_{in}^{T}\rangle+\langle\hat{\vec{X}}_{in}\hat{\vec{X}}^{T}+\hat{\vec{X}}\hat{\vec{X}}_{in}^{T}\rangle^{T}}{2}
\end{equation}
The last term of Eq. (\ref{App3}) is the diffusion matrix $ N $. Solving Eq. (\ref{Eq.8}), the quadratures can be written according to the input noises as:
\begin{equation}\label{App4}
\hat{\vec{X}}(t)=e^{Qt}\hat{\vec{X}}(0)+\int_{0}^{t}e^{Q(t-\acute{t})}\hat{\vec{X}}_{in}(\acute{t})
\end{equation}
Then the diffusion matrix will be given as:
\begin{equation}\label{App5}
N=\frac{1}{2}(A+A^{T})
\end{equation}
where
\begin{equation}\label{App6}
A=\int_{0}^{t} d\acute{t} \ [D(t,\acute{t})+D(\acute{t} ,t)^{T}] e^{Q^{T}(t-\acute{t})}
\end{equation}
where $ D(t,\acute{t}) = \langle \hat{\vec{X}}_{in}(t) \hat{\vec{X}}_{in}^{T}(\acute{t})\rangle $ and it is derived easily using the correlation functions of the input noises:
\begin{equation}\label{App7}
D(t,\acute{t})= \frac{\delta(t-\acute{t}) }{4}\ \left(\begin{array}{ccccc}
\bar{\gamma} & i \gamma & 0 & 0 & 0\\
-i \gamma &  \bar{\gamma} & 0 & 0 & 0\\
0 &  0 & \kappa & i \kappa & 0\\
0 & 0 & -i \kappa & \kappa & 0 \\
0 & 0 & 0 & 0 & 8 \gamma_{c}^{2} \Gamma_{L}
\end{array}  \right)
\end{equation}

where $ \bar{\gamma}=\gamma (1+2 N_{m}) $ and $ \bar{\kappa}=\kappa (1+2 N_{C}) $. The diffusion matrix $ N $ is obtained by employing Eqs. (\ref{App5}), (\ref{App6}) and (\ref{App7}):
\begin{equation}\label{App8}
N= \ \left(\begin{array}{ccccc}
\frac{\bar{\gamma}}{4} & 0 & 0 & 0 & 0\\
0 &  \frac{\bar{\gamma}}{4}  & 0 & 0 & 0\\
0 &  0 & \frac{\bar{\kappa}}{4} & 0 & 0\\
0 & 0 &0 & \frac{\bar{\kappa}}{4} & 0 \\
0 & 0 & 0 & 0 & 2 \gamma_{c}^{2} \Gamma_{L}
\end{array}  \right)
\end{equation}


\end{document}